# Evaluation of Game Templates to support Programming Activities in Schools

Bernadette Spieler[1], Christian Schindler[1], Wolfgang Slany[1], Olena Mashkina[1], María Eugenia Beltrán[2], Helen Boulton[3], David Brown[3]
[1]Graz University of Technology, Graz, Austria
[2]INMARK Europa, Madrid, Spain
[3]Nottingham Trent University, Nottingham, England
bernadette.spieler@ist.tugraz.at
christian.schindler@ist.tugraz.at
wolfgang.slany@tugraz.at
olena.mashkina@ist.tugraz.at
xllibel@gmail.com
helen.boulton@ntu.ac.uk
david.brown@ntu.ac.uk

**Abstract:** Game creation challenges in schools potentially provide engaging, goal-oriented, and interactive experiences in classes; thereby supporting the transfer of knowledge for learning in a fun and pedagogic manner. A key element of the ongoing European project No One Left Behind (NOLB) is to integrate a game-making teaching framework (GMTF) into the educational app Pocket Code. Pocket Code allows learners to create programs in a visual Lego®-style way to facilitate learning how to code at secondary high schools. The concept of the NOLB GMTF is based on principles of the Universal Design for Learning (UDL) model. Its focus lies on three pillars of learning: the *what*, *how*, and *why*. Thereby, the NOLB GMTF is a common set of concepts, practices, pedagogy, and methods. This framework provides a coherent approach to learning and teaching by integrating leisure oriented gaming methods into multi-discipline curricula. One output of this framework is the integration of game-based methods via game templates that refer to didactical scenarios that include a refined set of genres, assets, rules, challenges, and strategies. These templates allows: 1) teachers to start with a well-structured program, and 2) pupils to add content and adjust the code to integrate their own ideas. During the project game genres such as adventure, action, and quiz, as well as rewards or victory point mechanisms, have been embedded into different subjects, e.g., science, mathematics, and arts. The insights gained during the class hours were used to generate 13 game templates, which are integrated in Create@School (a new version of the Pocket Code app which targets schools). To test the efficiency of these templates, user experience (UX) tests were conducted during classes to compare games created by pupils who used templates and those who started to create a game from scratch. Preliminary results showed that these templates allow learners to focus on subject-relevant problem solving activities rather than on understanding the functionality of the app. This directly leads to more time to express their creativity in different levels and more time for extra tasks.

## 1. Introduction

Recent numbers gathered by the video game industry in Europe (Global Games Market Report 2016 and LAI Global Game Services, 2017), and selected statistics from all over the world (Entertainment Software Association, 2016 and Impos Connect, 2016) show that playing games is a popular leisure activity for the new generation of digital natives. Although pan-European initiatives (Balanskat and Engelhardt, 2014) try to shape school curricula more towards computer science and programming, pupils still leave school with a lack of computer science education. Often computer science is an optional course and is not equally distributed over the grades. For instance, in Austria, according to the Federal Ministry of Education's high school curriculum (Federal Ministry of Education Austria 1, 2017), computer science is mandatory only in Year 9 (Federal Ministry of Education Austria 2, 2017). From Year 10 to 12 it is optional (Federal Ministry of Education Austria 3, 2017) and therefore not all pupils are finishing school with meaningful Information and Communication Technologies (ICT) employability skills, which is a basic requirement for modern high-tech economies (Fraillon, et al, 2015). Spain has established compulsory content in the curriculum for the primary level to acquire information and digital competencies, whereas in secondary level computer science is optional for Year 8 (Gobierno de España Ministerio de Educació, 2009). Other European countries have a similar situation (Balanskat and Engelhardt, 2014). Thus, there exists an extreme pressure on schools to teach ICT, especially as the next generation of jobs will be char-acterised by an increased use of computer technology. Additionally, a German study (Renn, et al, 2009) shows the necessity for early contact with ICT topics to increase interest in such topics. Due to the underrepresentation of computer science (CS) topics in (high) school curricula, the amount of time to teach CS is rather limited and furthermore, today's teachers are seldom thoroughly trained in these topics. These factors impair the potential quality of engagement experienced by the pupils and teachers in these courses.

To achieve "social innovation" in the education sector, the NOLB project aimed to introduce a gamified approach (Kopcha, et al, 2016) into different pilot schools as part of their formal academic curriculum for delivering innovative teaching and learning practices. This gamified approach supports motivation and engagement thus makes learning more attractive and also addresses different behaviours such as collaboration, creativity, and self-guided studying (Caponetto, Earp and Ott, 2014). This paper focuses on the development of game templates as part of the NOLB GMTF. These learning materials should help teachers to guide their pupils through the practice of game development while learning about academic subject content at the same time.

The paper is structured as follows: Section 2 introduces the Pocket Code tool, which is used to directly support game creation on mobile devices. It describes the NOLB project aims, the design of the game-making teaching framework, as well as the incorporation of the Universal Design of Learning principles. Section 3 gives insights into the curriculum adaptation process, the teaching and learning process, the creation of the game-templates. Section 4 shows the outcome of the activities discussed in Section 3; the resulting game templates which are categorised by game genre and curriculum subject. In Section 5, the practical implementation of the different templates is described and evaluated. Section 6 contains the discussion and challenges of the chosen approach. The conclusion and a prospect of future work is described in Section 7.

## 2. NOLB Game-Making Teaching Framework

The NOLB game-making teaching framework (GMTF) aims at helping teachers to apply game mechanics, dynamics, academic content, and the Pocket Code app to the academic curricula. It provides a consistent structure of game elements, rules, and assets for the developed set of game templates. The framework is linked to the Universal Design for Learning (UDL) model and aims at motivating pupils and improving engagement by using game-based approaches in the curriculum. This chapter gives an overview of the Pocket Code app and the setup at the pilot schools within the NOLB project. The chapter concludes with the presentation of the adapted UDL principles, which overlap with the three pillars of the project's framework.

### 2.1 Pocket Code: A Programming Environment for Novices

Pocket Code is an Android based visual programming language environment built to allow the creation of games, stories, animations, and many types of other apps directly on phones or tablets. The interface provides a variety of pre-defined bricks that can be joined together to develop fully fledged programs. Complexity can vary from simple sequences of steps to create an animated story to the use of traditional programming concepts (branching, loops etc.) to create interactive games. The visual "lego"-style programming language used is very similar to that which is used in Scratch (https://scratch.mit.edu) and should support users in their first programming experience.

### 2.2 The No One Left Behind project: Experimental pilots

The ongoing NOLB project has the goal to adopt Pocket Code to academic curricula and to develop the GMTF. The project started in January 2015 and will end in June 2017. The NOLB project validates its output in three phases: feasibility study, first, and second cycle. The experimental pilots were conducted in Austria, the UK, and Spain, targeting 600 pupils between 8 and 18 years. The schools and teachers that agreed to participate in the project shared details about their lesson plans, curricula, and their ideas for projects. At the beginning of the project, teacher meetings and training sessions were conducted to ensure a smooth start of the study. For NOLB, pupils solved curriculum related problems while using the Pocket Code app. The teachers guided and assisted them in their learning processes.

### 2.3 Adapt Universal Design for Learning principles for NOLB

The Universal Design for Learning (UDL) supports varied learning and assessment approaches, such as cooperative learning, performance-based assessment, or pupil-centred learning (Spencer, 2011). The UDL applies advances in the understanding of how the brain processes information to the design of curricula, which accommodate diverse learning needs (CAST Inc, 2012). Therefore, the UDL provides a flexible approach that can be customised and adjusted for individual needs and supports the creation of teaching goals, strategies, and methods, as well as materials (Brand and Dalton, 2012).

For NOLB, these principles were linked to the GMTF, so that it responds to the basic questions of learning: *what* is being taught, *how* the information is shared, and *why* the information is engaging the learners. Based on these three UDL pillars, the GMTF refers to: curriculum adaptation and planning process, the teaching-learning process, and the assessment and feedback process, see Figure 1.

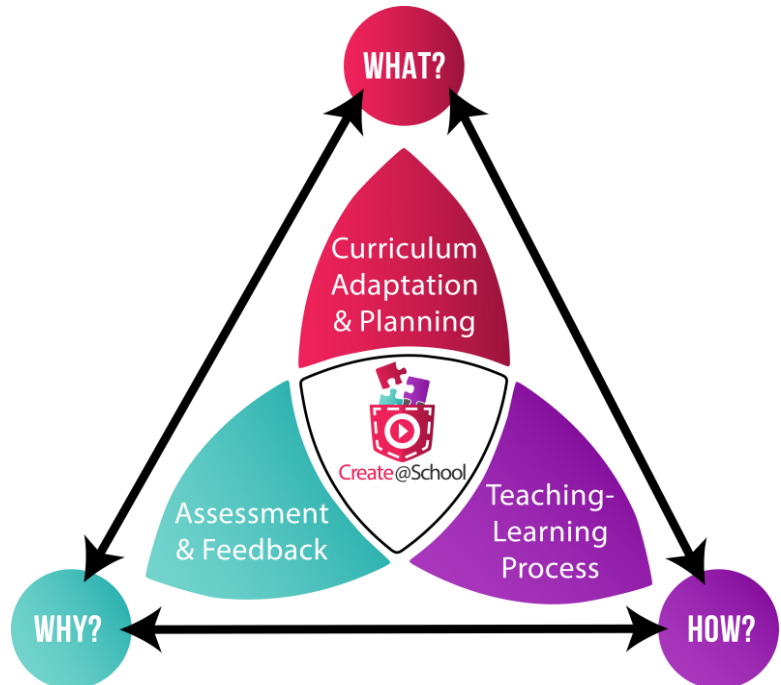

**Figure 1**: Link UDL principles to the NOLB GMTF

In practical terms, this framework offers a multidisciplinary approach to the:
- Creation and planning of game-based subject-relevant scenarios for templates (*what* they learn) e.g., physics, fine arts, music, etc.
- Review and integration of elements of game mechanics and dynamics into the teaching and learning process (*how* they learn)
- Assessment of relevant teaching and learning experiences (*why* they learn)

The components of each process of the GMTF are a collection of day-to-day and core practices, gathered from interviews and focus groups performed with teachers at the pilot schools, see Figure 2.

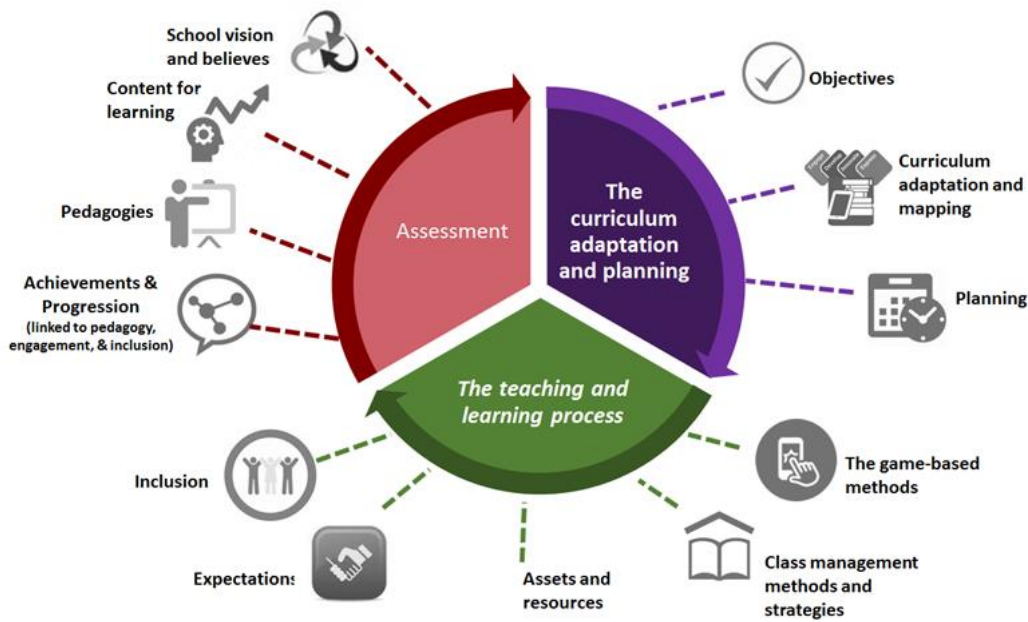

**Figure 2**: Components of the GMTF in NOLB

The curriculum adaptation and planning as well as the visualisation of the teaching and learning processes help to create the illustrative scenarios and to integrated game-based strategies. This is explained in more detail in the next section. The assessment and feedback process is part of Section 5.

## 3. Creation of the Game Templates

By merging the *what* and *how* of learning, the result will be the following: a) clear objectives for each class and subject, and b) standardised game-based methods, including game mechanics and dynamics. This chapter explains how both influenced the creation of the game templates. Therefore, it was necessary that the developed game templates were applicable to several subjects and classroom levels, and universally recognised as "games" by pupils who are familiar with best practice game examples. To link game mechanics and game dynamics with academic content, Pocket Codes' functionality has been adopted, which resulted in the new Create@School app for academic purposes. This is described at the end of the section.

### 3.1 Curriculum Adaptation and Planning process

For NOLB, this process helps 1) to define class objectives, 2) curriculum adaptation and mapping, and 3) the planning process, see Figure 3. The class objectives are based on learning goals and take into account non academic and cognitive skills, as well as computational thinking skills (Walden, et al, 2013). Therefore, the learning goals of each game must be clearly defined and build towards the overall objective of the class (Clapper, 2009). For curriculum adaptation, the objectives of selected subjects were prioritised and this curriculum was mapped in the developed game templates. Thus, it gives a clear line of sight between academic content to be delivered, workload to be given, and individual needs. The planning process includes the timeline for the lessons required for acquiring the needed skills completing the tasks of the templates to be assigned by using Create@School.

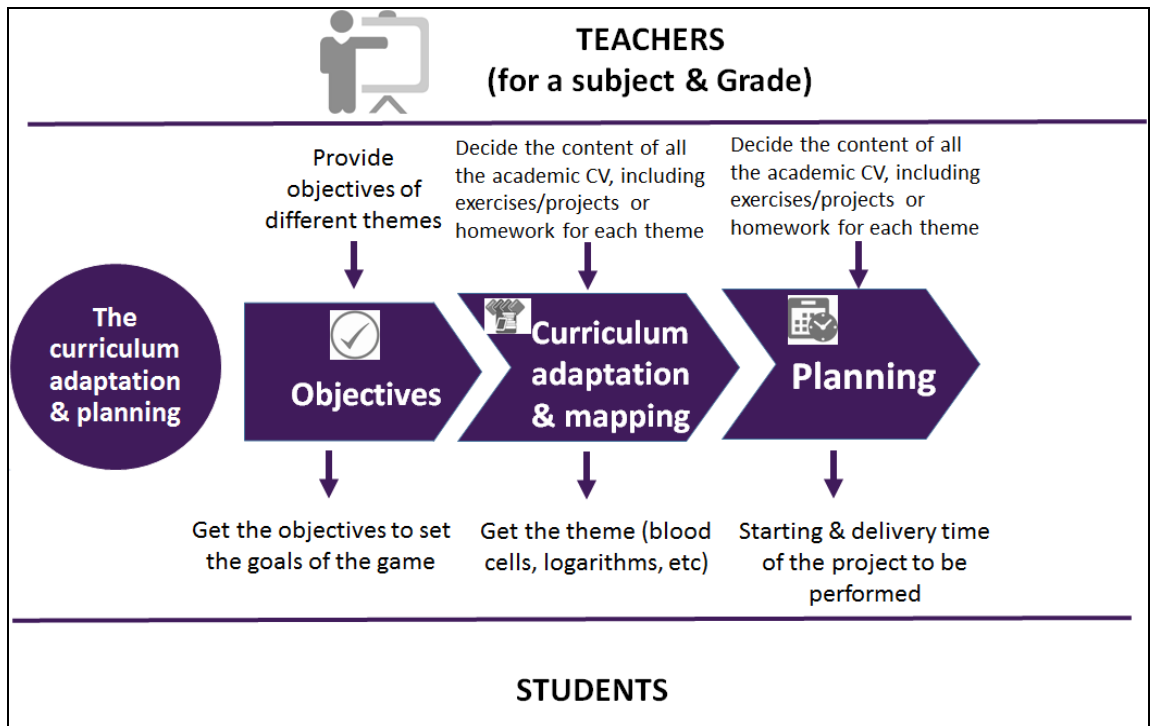

**Figure 3**: The curriculum adaptation and planning component

### 3.2 The Teaching and Learning process

This process comprises the integration of game-based methods, assets, and strategies to be followed in classes. In this phase game elements, game genres, and learning assets and resources to be used for the game's templates were defined.

#### 3.2.1 Mechanics, Dynamics, and Aesthetics

Game mechanics, dynamics, and aesthetics (MDA) provide a consistent structure to define game elements, goals and rules and thereby deliver a common framework and vocabulary for games (Hunicke, LeBlanc and Zubek, 2004). The MDA is a formal approach to understand games and their elements in order to support the process of designing and developing a game (Bohyun, 2015). *Mechanics* are a synonym for the "rules" of the game and defines the objects, elements, and their relationships. They comprise points, levels, challenges or virtual goods. *Dynamics* describe the play of the game when the rules are set in motion. They comprise rewards, status, achievement, self-expression, or competition. *Aesthetics* refer the player's experience with the game. They are the reason for playing games and comprise, for example, fantasy, narrative, fellowship, and discovery.

#### 3.2.2 The "Shape of a Game"

To use the same shape[1] for all game templates, the following structure has been developed:
- Title screen
- Instructions screen: Conveys "goal" and "rules".
- One or more levels: Use of the word "level" creates a connection to commercial games.
- An end screen: This is linked to the end of a story, the achievement of a target, etc.

#### 3.2.3 Clustering of Game Genres

Video games exist largely in a commercial entertainment marketplace, and have evolved around clusters (Ferreira, et al, 2008). The classifications of game genre are unfixed and diverse. The genres in Figure 4 have been identified as offering useful models for gameplay (Lee, et al, 2014).

---

[1] The project drew upon the work of partners' visiting speaker Gary Penn, formerly Creative Director at DMA Design (maker of the original Grand Theft Auto game).

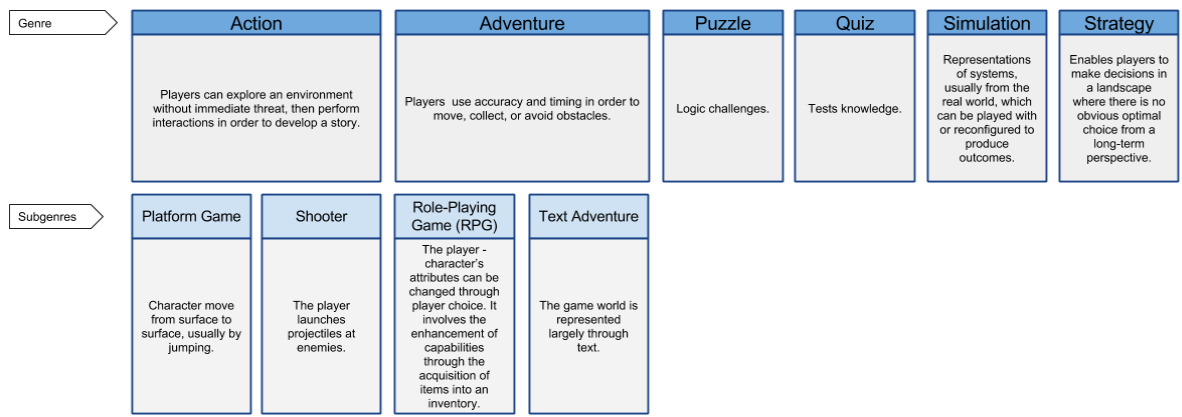

**Figure 4:** Clustering of game genres in NOLB

The game genres helped to define what game design elements are necessary to effectively create the chosen genre and that the theme 'fit' into the genre classification. By adopting learning content into something that appears to be a game, a new experience for pupils is created.

### 3.3 The Create@School app: Adopt Pocket Code to academic standards

Based on the results of the first cycle a more enhanced version of the Pocket Code app for schools was developed. Pocket Code helped its users to learn basic programming steps but the app was difficult to transfer to other subjects besides computer science. The app needed improvement in terms of the usability by decreasing both complexity and the time needed to develop games. Therefore, a more powerful and usable interface for the management of objects and large programs has been integrated into Create@School. For example, the scene feature was integrated for creation of large programs and management of different levels. An example of scenes overview is shown in the third image from the left in Figure 5. The new app, named Create@School, was released on October 2016 for use at the pilot schools.

### 4. Results

This Chapter provides an overview of the game templates, see Table 1. They are all based on the "Shape of a Game" and were translated in the languages of the participated countries. Template tutorials for teachers and pupils were created (see the project's edu-platform[2]). A new option "Templates" has been added to the Create School's main menu to access the templates list, see Figure 5.

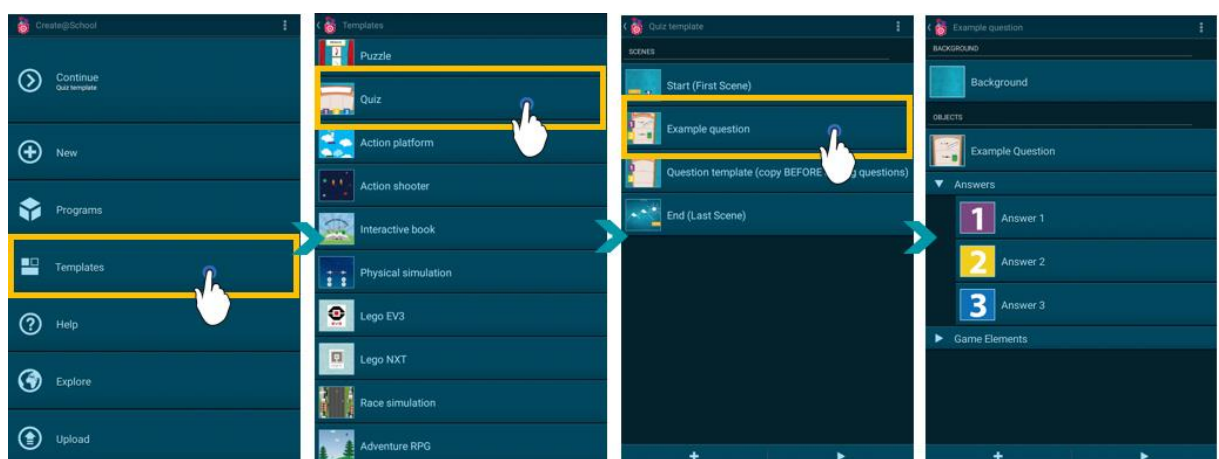

**Figure 5**: Game templates in Create@School that supports scenes and grouping of objects

By changing or adding different contexts, game assets, or game mechanics the pupils adapt, customise, and create diversity with the dynamics and aesthetics of the games. By using the game design elements they can build new games and remix existing ones. The templates allow to edit the existing design, giving the pupils freedom in the personalisation of backgrounds and characters. Fun, engaging experiences are generated through the creation of new, challenging levels or changing the difficulty of a game.

---

[2] https://edu.catrob.at/

**Table 1:** Overview of the developed game templates

| Genre | Subject | Theme | Learning goal | MDA |
|---|---|---|---|---|
| *Quiz* | Physics | Properties of physical objects | Learn about physical objects and their properties (e.g., inertia) through questions and answers. | points, knowledge, narrative |
| *Adventure* | Science | Space | Listen to a space scenario and decide "Yes" or "No". | levels, discovery, narrative |
| *Puzzle* | Music | Instrument groups | Tap on the musical instrument, which does not fit to the group (odd one out). | points, timer, logic rules, narrative |
| *Action* | Science | Respiration | Learn about de-/oxygenated blood cells by tapping on the objects. Avoid the virus cells. | points, timer, high-score, fellowship |
| *Action Platform* | Physics | Periodic system | Move the character from surface to surface, to catch halogens. | Points, timer, narrative |
| *Physical Simulation* | Physics | Newton's laws of motion | Perform physical experiments with Newton's 2nd law of motion and experiments with the formula. | Submission, expression, discovery |
| *Action Shooter* | Maths | Division rules | Shoot asteroids which can be divided by 4, 3 and 11. | levels, points, timer, narrative |
| *LEGO Simulation* | Computer Science | Robotics | Use the LEGO NXT / EV3 extension and solve tasks (e.g. creating a maze). Learn about sensor values, or coordinates. | levels, submission, challenge |
| *Interactive book* | Science | Water cycle | An easy version of an interactive book with scenes that explains the water cycle. | levels, submission, expression |
| *Adventure RPG* | Fine Arts | Colour circle | The user has to collect colours (inventory) to draw a picture. Additional: Create your own character. | levels, virtual good, achievement, challenge |
| *Racing simulation* | Science | Pollution | Collect trash to get points and level up. Avoid the other cars to keep playing. | levels, points, challenge, achievement |
| *Life simulation* | Life after school | Work skills | Help the character serve the right order and keep the customers happy. | levels, points, challenge, achievement |
| *Strategy* | History | *not finished yet* | Multiplayer (on one tablet). Strategy decisions via a simple connect 4 template. | Points, leadership, narrative |

## 5. Assessment and Feedback: Evaluation of the game templates

Assessment and feedback (the *why*) is provided through the visualisation of analytic data in the Project Management Dashboard (PMD) including both qualitative and quantitative data. Through the integration of a BDSClientSDK (Big Data Services Client software development kit) in Create@School it is possible to explore information about users, their sessions, and actions. The PMD provides the framework needed for teachers to manage pupils and their projects. Explaining both assessment tools in more detail would be beyond the scope of this paper.

The analytics data that refers to the templates and results of the UX tests are presented within this section, as the focus of this paper lies on the created templates. This data was collected between February and April 2017. All in all 378 pupils (67 in Austria, 154 in Spain, and 157 in UK) took part in the project during this time period, see Table 2. The tracked data show that the event with the name "useTemplate" was logged 183 times. This means 183 games were created on the basis of a template. Further, the event with the name "createProgram" was logged 233 times in all pilot schools, which means these programs were created without the use of a template. The setup of the school units in which the templates has been used was very similar through all countries. Therefore the lesson plans' class teaching strategy was broken into three main sections: To start with (starter), Main Learning and Extension (plenary). The starter includes a debug or basic coding activity to support the learning, but it is the main learning section where pupils start coding, predominantly independently.

**Table 2:** Results of the experiments presented on countries, ages and themes

| Country | Age | Pupils | Units | Template | Theme | Learning goal |
|---|---|---|---|---|---|---|
| UK | Yr 5 (9/10 year old) | 30 | 3 | Adventure | Science / Space | to apply their space themed written work in literacy to adventure games. |
| UK | Yr 8 (12/13 year old) | 27 | 2 | Quiz | Science / Energy | to demonstrate and assess their understanding of the unit of work as a whole. |
| UK | Yr 8 (12/13 year old) | 27 | 2 | Action Platform | Science | to collect halogens and avoid non-halogens. |
| Spain | Yr 8 (12/13 year old) | 9 | 4 | Interactive Book | Science / Biology | to create an interactive book on a vegetable garden. |
| Spain | Yr 9 (13/14 year old) | 22 | 4 | Interactive Book | Maths | to learn about logic and arithmetic. |
| Austria | Yr 12 (17/18 year old) | 12 | 4 | Quiz | Computer science | to create questions about the history of their school. |
| Austria | Yr 8 (12/13 year old) | 25 | 4 | Physical Simulation | Physics / Motion | to expand the template and add an own rocket. |

This paper evaluates the use of the Physics Simulation template in more detail to test the efficiency of the templates and to compare games created based on a template and without it. Therefore, the data of two classes in Austria were compared. Both times the app was used in Physics in Year 8 (12/13 year old) with the same teacher and the same amount of pupils (25 pupils). The lesson with the first class was conducted during the feasibility study (December 2015) and the second class had their lessons during the second cycle (March 2017). After both units a post-questionnaire was conducted with pupils and an interview with the teacher was performed, see the results in the Tables 3 and 4.

**Table 3:** Physics project with and without the use of a game template

|  | **First class (December 2015)** | **Second class (March 2017)** |
|---|---|---|
| *Number of units* | 6 units à 45 minutes:<br>• 4 starter units<br>• 2 programming units | 4 units à 45 minutes:<br>• 2 starter units<br>• 2 programming units |
| *Theme* | Density of objects and liquids<br><br>Formula:<br>object density < fluid density = item floats<br>object density > fluid density = item sinks | Newton's' second law of motion<br><br>Formula:<br>force = mass * acceleration |
| *Learning goal* | • Add an object and let it glide<br>• Set/Show variables to define the properties of the objects<br>• Apply the formula | |
| *Starter unit* | Every pupil starts with a new program. One adds one step to the program in front of the class (this was displayed on the projector). At the end of the instruction units every pupil had one example level integrated for reference. | The class programs one game together. The finished program was a simple "ping-pong" game which used physical properties. In the second unit everyone used their own devices and played the first example level, adjusted the code, and answered the question after applying the formula. |
| *Main Learning and Extension units* | Add a new object (own drawing) and define its movement and behaviour by adjusting its density. Apply the formula. | Add own code and a look (picture of self-constructed rocket) to the empty object within the template level and define its behaviour. Apply the formula. Extra level for pupils who finished level 2 earlier (Newton's 3rd law of motion). |
| *Learner achievement* | **6 of 25 reached the learning goal** | **16 of 24 reached the learning goal**<br>(one was absent during programming unit) |
| *Summary teacher interview* | The teacher commented that the project was a little bit too difficult for the pupils. Some of the pupils had more problems doing the density program and some were quite fast. | The teacher said that the project was very successful. She found it nice to see the pupils engaged and working on their program. |

**Table 1:** Summary of the student's surveys

| Questions | First class (December 2015) | Second class (March 2017) |
|---|---|---|
| *How did you like the experience with the app?* | 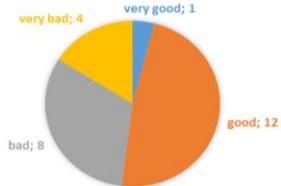 very good; 1<br>good; 12<br>bad; 8<br>very bad; 4 | 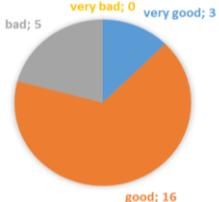 very good; 3<br>good; 16<br>bad; 5<br>very bad; 0 |
| *What did you like the most?* | They liked that they could make their own apps, and that they learnt something about coding. | Most frequently mentioned: It was very straightforward and simple, the programming of the rocket itself was fun, and pupils liked the possibility to add their own pictures. |
| *What did you like the least? Any suggestions for app improvement?* | They mostly mentioned that it was confusing at the beginning and that it was very time consuming to have the first simple steps.<br><br>Improvements: more help functions, tutorials, more tools, less typing. | The most mentioned comment was that the game to create was still seen as too complicated and it was confusing them.<br><br>Improvements: To add a search function for bricks. |
| *Have you used Pocket Code in your spare time?* | None of them.<br><br>Reasons for not using Pocket Code: no interest, no time, too difficult. | 2 used it in their spare time.<br><br>Reasons for not using Create@School: no interest and no motivation to use it. |

## 6. Discussion

Based on the results from 2015, not only the program but also the whole packaging of the course was adapted. For example, the pupils received a more general starter lesson about the app itself. Therefore, the template integrated the idea to let pupils first change the existing code, thus understanding the overall concept of the game template. In the second step, they had to add a similar object and apply the same concept they learnt from the previous level. Therefore, instead of four instruction units, only two were needed. The on-site observations showed that the second class understood how to apply the physical formula better and solved the physics related problem. In contrast, the first class dealt more with programming/app problems and most of them did not reach the point to apply the physical formula. Thus the pupils of the second class felt more engaged, were more concentrated, and more of them reached the predefined learning goal as a result. It addition, it was seen as very positive that the template allowed personalization by adding pupil's own picture of their self-constructed rockets.

In general, all templates were seen as easy to understand, use and expand. For example, for the quiz template in both countries, pupils needed time to integrate the first question, but after they understood the overall concept, they added the other questions very quickly.

The challenge for developing the templates was, on the one hand, to pre-program the templates in an efficient way which provides pupils with functionalities to aid the process of building a new game; the templates must explain the important dynamics, mechanics, and aesthetics of a game (like coding how to reward the achievements or how to collect points). On the other hand, pupils should have the freedom to express themselves in a creative, fun, and dynamic way (e.g., to change images, sounds). Every class project should show them different methods to achieve the goals in a way which supports their logical thinking processes.

## 7. Conclusion and Future Work

The research for this paper is dedicated to provide appropriate teaching materials in the form of predefined templates to teachers and therefore to find new possibilities for adopting a gamified approach in schools. The results showed that the developed game templates encourage learning by doing, allow the expression of one's own ideas, and provide a visual programming language that is easy to understand and to learn. The project's goals were archived by linking the generated game templates to the program and design patterns of commercial games and thereby effectively support the development and adaptation of the learning material in a structured and replicable manner.

The work to be completed in the future is to estimate the effectiveness of the remaining templates and eliminate the factors that can bias the results. One important point for the further evaluations is if the templates can be applied to different subjects. For this purpose several classes will use the templates from May to June 2017. A usability test is planned with the Adventure Interactive Book template (Year 8, 12/13 year old), and the Adventure RPG (Year 10, 14/15 year old). For the purpose of evaluation, interviews and surveys about the templates will be performed, opposed to only asking the pupils questions about the app. To define the general applicability of this template, the learning goal for computer science classes will be to apply the template to one randomly allocated subject (e.g., biology or history).

## Acknowledgements

This work has been partially funded by the EC H2020 Innovation Action No One Left Behind, Grant Agreement No. 645215.

BibTex entry:

```
@InProceedings{SPIELER2017ECGBL,
author = {Spieler, B. and Schindler, C. and Slany, W. and Mashkina, O. and Beltrán M.E. and Boulton H. and Brown D.},
title = {Evaluation of Game Templates to support Programming Activities in Schools},
series = {In Proceedings of the 11th European Conference on Game Based Learning},
booktitle = {Proceedings of the 11th European Conference on Game Based Learning},
isbn = {9781510850446},
issn = {2049-0992},
publisher = {Academic Conferences and Publishing International Limited, 2017},
location = {Graz, Austria},
month = {5-6 Oktober, 2017},
year = {2017},
pages = {600-609}}
```